# Enabling Technologies for Visible Adaptive Optics: The Magellan Adaptive Secondary VisAO Camera

Derek Kopon[*a], Jared Males[a], Laird M. Close[a], Victor Gasho[a]
[a]CAAO, Steward Observatory, University of Arizona, Tucson AZ USA 85721

## ABSTRACT

Since its beginnings, diffraction-limited ground-based adaptive optics (AO) imaging has been limited to wavelengths in the near IR ($\lambda > 1\mu m$) and longer. Visible AO ($\lambda < 1\mu m$) has proven to be difficult because shorter wavelengths require wavefront correction on very short spatial and temporal scales. The pupil must be sampled very finely, which requires dense actuator spacing and fine wavefront sampling with large dynamic range. In addition, atmospheric dispersion is much more significant in the visible than in the near-IR. Imaging over a broad visible band requires a very good Atmospheric Dispersion Corrector (ADC). Even with these technologies, our AO simulations using the CAOS code, combined with the optical and site parameters for the 6.5m Magellan telescope, demonstrate a large temporal variability of visible ($\lambda=0.7\mu m$) Strehl on timescales of 50 ms. Over several hundred milliseconds, the visible Strehl can be as high at 50% and as low as 10%. Taking advantage of periods of high Strehl requires either the ability to read out the CCD very fast, thereby introducing significant amounts of read-noise, or the use of a fast asynchronous shutter that can block the low-Strehl light. Our Magellan VisAO camera will use an advanced ADC, a high-speed shutter, and our 585 actuator adaptive secondary to achieve broadband (0.5-1.0 µm) diffraction limited images on the 6.5m Magellan Clay telescope in Chile at Las Campanas Observatory. These will be the sharpest and deepest visible direct images taken to date with a resolution of 17 mas, a factor of 2.7 better than the diffraction limit of the Hubble Space Telescope.

## 1. INTRODUCTION: MAGELLAN VISIBLE ADAPTIVE OPTICS

The Magellan Clay telescope is a 6.5m Gregorian telescope located in Chile at Las Campanas Observatory (LCO). The Gregorian design allows for a concave F/16 adaptive secondary mirror (ASM) that can be tested off-sky with a retro-reflecting optic at the far ellipsoidal conjugate. We have fabricated an 85 cm diameter adaptive secondary with our subcontractors and partners that uses 585 actuators with <1 msec response times and will allow us to perform low emissivity AO science. We will achieve very high Strehls (~98%) in the Mid-IR AO (8-26 microns) that will allow the first "super-resolution" and nulling Mid-IR studies of dusty southern objects. We will employ a high order (585 mode) pyramid wavefront sensor (WFS) similar to that used in the Large Binocular Telescope AO systems. (For more on the Magellan ASM and mid-IR AO, see Close et al. 2008)

The AO system currently being built for the Magellan telescope (Figure 1) consists of an ASM built by the University of Arizona mirror lab and a pyramid wavefront sensor (WFS) built by the Osservatoria Astrofisico di Arcetri (Esposito et al. 2008). The ASM is identical in optical prescription to the LBT ASMs and will use all the same hardware and control software (for more on the LBT AO system, see Riccardi et al. 2008). The primary infrared science camera is BLINC/MIRAC4 (Hinz et al. 2009), which will receive IR light from a dichroic beam splitter. Visible light reflected by the dichroic will be sent to an optical bench (hereafter called the "W-Unit") containing the WFS and a visible (0.5-1.0µm) science CCD. The layout of the W-unit is shown in Figure 3.

The relatively high actuator count of our ASM will also allow us to obtain modest Strehls in the visible (0.5-1.0 µm). Our VisAO CCD47 camera is fed by a beamsplitter piggybacked on the W-Unit. Taking full advantage of the periods in a night when the seeing is 0.5" or better requires the Vis AO camera to "always be ready". Because the Vis AO camera is conveniently integrated into the WFS stage, we can select a beam splitter to steer a percentage of the WFS visible light into the Vis AO camera with 8.5 mas pixels. While our ASM and WFS are essentially the same as those on the LBT, the VisAO camera has a number of novel design features for optimizing visible science.

The excellent seeing conditions at the Magellan site frequently provide $r_o$ as high as 20 cm at 0.55 µm. Because of this, we expect that at $\lambda \sim 0.9$ µm there will be AO correction on bright stars and that moderate Strehls will be possible in the I and z bands on good nights. The resulting angular resolution will be a spectacular 20-30 mas (although the corrected FOV will typically be limited by the isoplanatic angle to less than 8.5"). Our optical Strehl estimates (Esposito et al. 2008) may be optimistic (as these AO Strehl estimates tend to be). However, there is reason to expect that in the z, I, and R bands Strehls will be at least greater than 15%, 10%, and 5%, respectively, when $r_o$>20 cm for stars brighter than V=10. While these Strehls are low compared to what will be achieved at 10 μm, there is still a large body of science that can be done at low Strehl. Most current ~200 actuator 8-10m AO systems do not achieve Strehls much higher than 2% in the I band (0.85 μm). If we estimate no better control than these current systems, and note that our fitting error is a factor of 2x $rad^2$ better, then it is clear that our Strehls with bright stars (fitting error limited) will trend towards I band Strehls of 16%. Our objective for the CCD47 is diffraction limited image quality over the full 8.6" FOV over the band 0.5-1.0 μm out to a Zenith angle of 70°.

---

[*] dkopon@as.arizona.edu

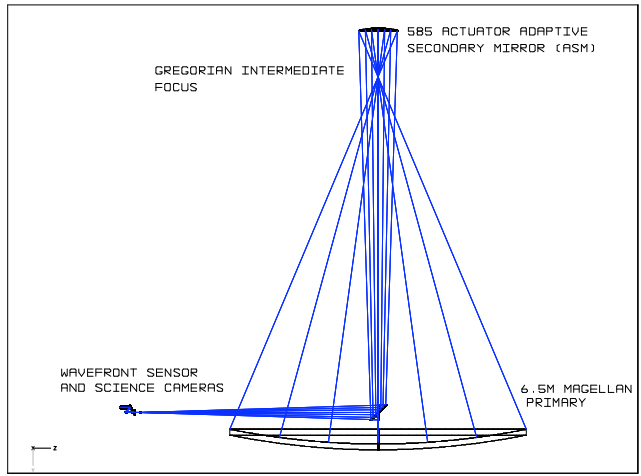

**Figure 1:** Ray trace of the 6.5m Magellan Telescope with the F/16 adaptive secondary. Note the Gregorian intermediate focus.

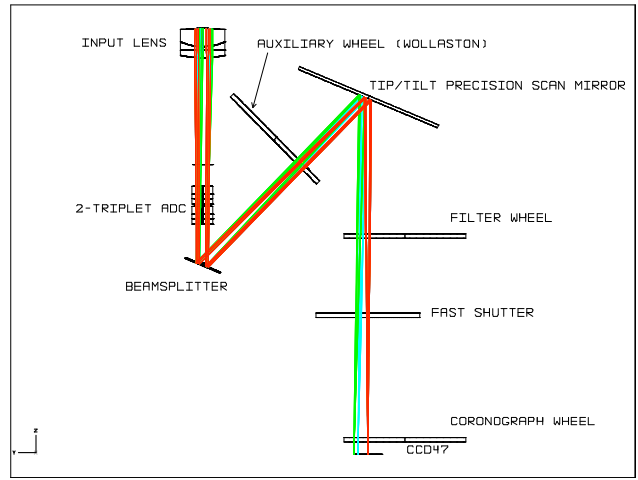

**Figure 2:** Ray trace of the VisAO science channel on the W-unit.

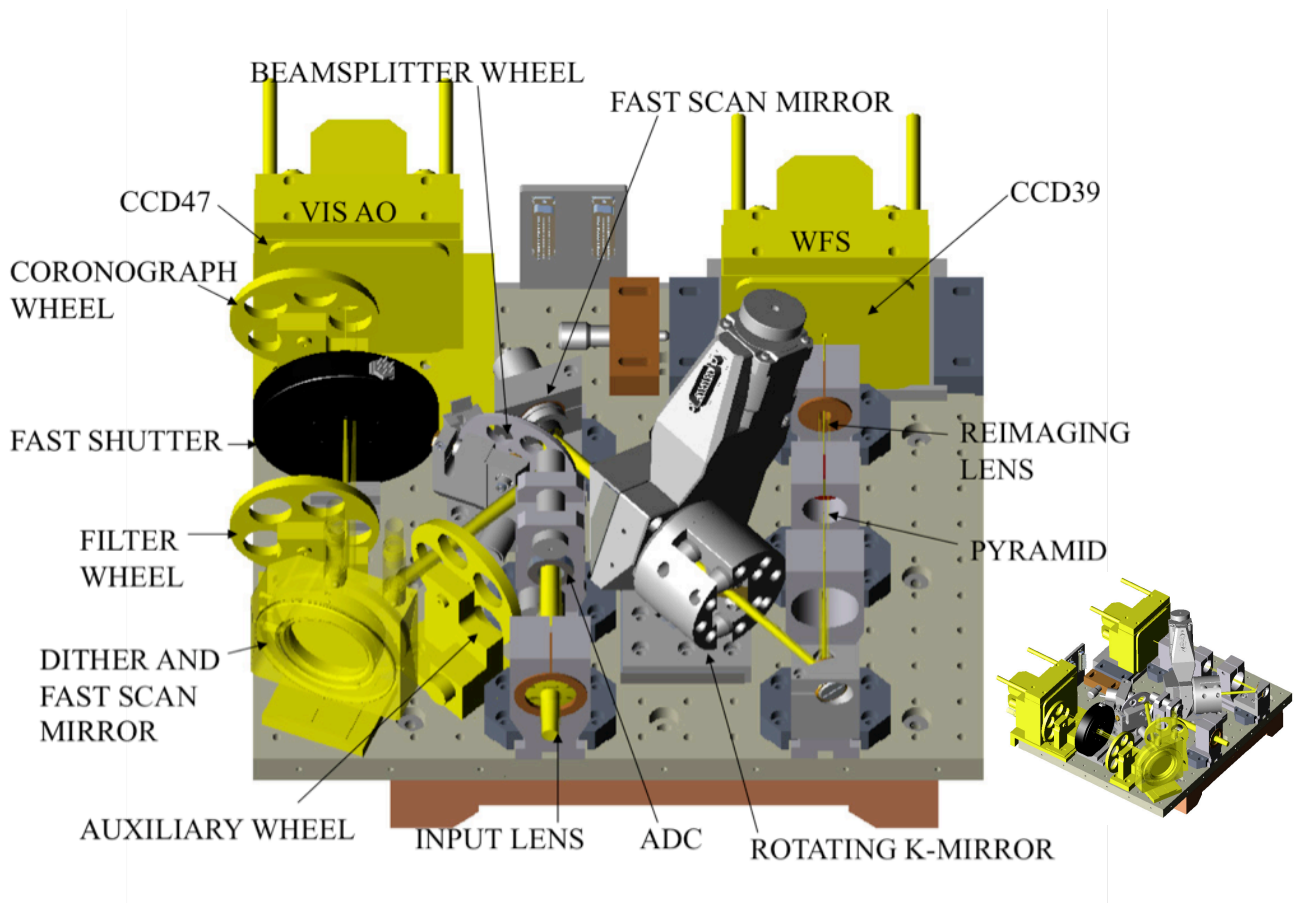

**Figure 3:** Layout of the W-unit. The beamsplitter sends reflected light to the VisAO science camera and transmitted light to the pyramid wavefront sensor.

Magellan is 2.7 times larger than HST and therefore will make images 2.7x sharper at the same λ. However, all the existing (operational) cameras on HST do not Nyquist sample wavelengths less than 1μm (it should make a ~48mas FWHM image at 0.6μm). In other words our VisAO camera makes images >4.7x better than HST in terms of pixel resolution. While drizzle can help HST, in general HST will be limited to resolution of ~60 mas at best (1.5 pixels FWHM limit to drizzle restoration). On the other hand, we should achieve images with FWHM=18mas at 0.6 μm (although at much lower Strehl). Therefore, we expect that our VisAO camera will make images ~2.7-3.3x sharper in FWHM than HST from 0.6-1.0 μm (at the same λ for each) due to our much smaller pixels and larger aperture.

Using our Magellan telescope parameters and atmospheric site surveys of the Las Campanas site, we use the CAOS code to simulate the performance of our AO system. In addition to determining an estimate of our required guide star magnitude, we have calculated an estimate of the Strehl ratio (SR) variability of our system at visible wavelengths. The high variability of visible Strehl, fluctuating between values as high as 50% and as low as 10%, has led us to design a fast asynchronous shutter that will use information from the wavefront sensor to block the low Strehl light. Using this shutter, we will obtain the benefit of being able to select periods of higher Strehl, without taking the corresponding read-noise hit.

## 2. THE W-UNIT: PYRAMID WAVEFRONT SENSOR AND VISAO SCIENCE CAMERA

The W-Unit (Figures 2 and 3) is an optical board located on three translation stages that can patrol a 2.3x3.2 arcmin field at the Nasmyth focal plane in order to acquire NGS guide stars and VisAO science targets. The W-Unit contains two optical channels: the pyramid wavefront sensor channel and the VisAO CCD47 science channel. Incoming visible light passes through a telecentric lens and a triplet lens that converts it from a diverging F/16 beam into a converging F/49 beam. This light then passes through the ADC before hitting a beam splitter wheel. Light transmitted by the beamsplitter goes to the WFS and reflected light goes to the VisAO science camera.

### 2.1 Pyramid Wavefront Sensor Channel

The WFS channel consists of a fast steering mirror, a K-mirror rerotator, a double pyramid, a reimaging lens, and the CCD39. The resultant image on the CCD39 is four pupil images whose intensity variations can be used to determine the wavefront. A detailed description of the operation of the pyramid sensor (PS) arm of the W-unit can be found in Esposito et al. 2008. The pyramid sensor is very important for visible AO because of its potential for diffraction limited performance and variable sensitivity. A Shack-Hartmann (SH) sensor is diffraction limited by the size of a pupil sub-aperture: i.e. the pitch of the lenslet array. The PS uses the full pupil aperture and is only diffraction limited by the size of the primary mirror. Since the wavefront sensing wavelengths are essentially the same as the science wavelengths (~0.7 μm), it is essential that the wavefront sensor be as close to the diffraction limit as possible (Esposito et al. 2000).

Another advantage of the PS that makes it crucial for visible AO is the dynamic range provided by the modulation of the piezo fast steering mirror. The fast steering mirror is located very close to the pupil and is used to scan the image of the guide star around the tip of the pyramid. The amplitude of this scan can be adjusted to a large value for a highly aberrated wavefront and a smaller value for a partially corrected wavefront. As wavefront correction becomes better, the amplitude of the scan can be decreased in order to increase sensitivity. This is in contrast to a SH sensor, whose sensitivity is essentially fixed by the lenslet pitch. The dynamic range provided by the PS should allow us to use ~2 mag fainter guide stars than would be allowed by the SH sensor, in addition to allowing us to come very close to diffraction limited correction (Esposito et al. 2001).

## 2.2 VisAO CCD47 Science Channel

The light reflected from the beam splitter will travel to the CCD47 VisAO camera, which will be used as both an acquisition and visible science camera. This light passes through an auxiliary wheel that contains various components, such as a grism, Whollaston prisms, etc. (see figure 4). It then reflects off of a remotely actuated fine adjustment/dither tip/tilt mirror that can be used to steer the guide star onto the coronagraphic spots at the focal plane. The light then passes through the filter wheel, the high-speed shutter, and the coronagraphic spot wheel before landing on the CCD47. The converging F/49 beam results in a square FOV of 8.6". Our goal for the CCD47 is diffraction limited image quality over the full 8.6" FOV over the band 0.5-1.0μm out to a Zenith angle of 70°. Meeting this tight performance requirement over a broad band at high zenith angles requires a high performance ADC.

## 2.3 W-Unit Components (ADI, Grism, coronagraphic spots)

The beamsplitter wheel will have various beamsplitters and dichroics that can be chosen based on the science target and the brightness of the guide star. See Figure 4 for a list of components in the various wheels – in order from farthest to closest to the CCD47.

The auxiliary wheel near the pupil (the "pupil" wheel in Fig. 4) will contain several components for various science modes of operation. A grism will allow slitless spectra of point sources to be taken. Wollaston prisms will allow SDI Hα imaging by splitting the beam into two beams with opposite polarizations that are imaged on different halves of the focal plane. Detection of scattered (hence linearly polarized) light of debris disks (and young circumstellar disks) will be possible. All four stokes parameters will be measured with the $0^o$ and $45^o$ Wollastons.

Because any bright sources will immediately saturate the Vis AO camera and potentially cause blooming, we will have a 0.1" chrome dot and a 1.0" chrome dot on selectable plates immediately in front of the CCD47 that will act as neutral density filter "coronagraphs". In order to steer the star onto one of these dots, or to look at sources with the CCD47 that are off-axis from the guide star being imaged on the pyramid sensor, the fold mirror immediately prior to the wide-field lens location will be an automated fine-adjustment tip/tilt stage. The wheel will also contain a split component with two coronagraphic "bars" for use with the SDI device.

| Pos# | Beam splitter R/T 0.5" dia | Pupil wheel 1inch dia | Filter wheel 1inch dia | Coronagraph wheel 17x17mm |
|---|---|---|---|---|
| 1 | Dichroic IRV/z | open | open | 1.0" dia chrome "bar"[*] |
| 2 | Mirror 96/4% | Woll 0[*] | z: 0.85-1.2 μm | 0.1" dia chrome spot |
| 3 | Dichroic z/IRV | Woll 45[*] | i: 0.7-0.85 μm | 1.0" dia chrome spot |
| 4 | AR Glass 1/99% | 90% | r: 0.55-0.69 μm | SDI $zH_20$, 1" "bar"[*] |
| 5 | ND3 2/0.1% | Grism | TBD | SDI Hα[*] |

**Figure 4:** Components for the beamsplitter wheel, the two VisAO filter wheels, and the VisAO coronagraph wheel

## 2.4 Ghosting Analysis

All of the refractive optics in the VisAO arm of the W-unit are potential sources of ghosts. This includes the input lens, the ADC, the beamsplitter, the filters, and the coronagraphic spots. The input lens, the ADC, and the coronagraphic spots will be AR coated to better than R < 0.01. In addition, ghosts originating in the ADC or the input lens will be out of focus by 2x the thickness of the element, which for both elements is on the order of ~20mm. Hence, any ghosts will not be focused images. Additionally, our Zemax analysis shows that our 1.0" coronagraphic spot will block all bright ghosts.

We performed a full ghosting evaluation of the W-unit optics in the CCD47 optical channel at 750 nm wavelength using the Zemax ghost focus generator. The ghost focus generator calculates the double and single-bounce ghosts produced by light reflected off of refractive elements. Our analysis looked at all double bounce ghosts off of all refractive elements surfaces and interfaces. For each potential ghost a new Zemax model was created that could be used to generate spot diagrams, ghost location, F/#, etc. In order of first to last in the optical path, the relevant refractive optics are the telecentric lens, the input lens, the two ADC elements, the flat piece of glass that contains the coronagraphic spots, and the first surface of the CCD. Other refractive optics, such as the grism or the filters were not considered, since these elements can be tilted easily so that the ghost does not fall on the detector. Ghosts whose first or second bounce included one of the tilted surfaces in the ADC where not considered, since these ghosts are also reflected well away from the detector. In summary, all "in-focus" ghosts will be hidden behind our coronagraphic spot while out-of-focus ghosts larger than our 1.0" spot will be >10,000x fainter than the halo level past the spot --and so they will be too faint to be a significant source of noise.

The filter wheel will be tilted sufficiently to move any filter ghosts far outside of the FOV. A wedge could be used to mitigate the ghost from the beamsplitter. However, past experience has shown that a focused, well-characterized beamsplitter ghost can be useful for astrometry and photometry since the central bright star will be obscured by the coronagraphic spot. Having a beamsplitter ghost of the central bright target star near the edge of the VisAO FOV is somewhat analogous to using a $10^5$ neutral density filter. This off-axis source can then be used as a tool of accurate astrometry on the central star when it is saturated (or behind the coronagraphic mask). The coronagraphic spot will be chrome deposited on the side of the glass facing the CCD47.

## 3. THE ADVANCED ADC

In order to achieve diffraction limited performance over this broad band, 2000μm of lateral color must be corrected to better than 10μm. The traditional atmospheric dispersion corrector (ADC) consists of two identical counter-rotating cemented doublet prisms that correct primary chromatic aberration. We propose two new ADC designs: the first consisting of two identical counter-rotating prism triplets, and the second consisting of two pairs of cemented counter-rotating prism doublets that use both normal dispersion and anomalous dispersion glass in order to correct both primary and secondary chromatic aberration. At high Zenith angles, the two designs perform 58% and 68%, respectively, better than the traditional two-doublet design. The ADC also serves to increase the sensitivity of the PS by allowing smaller scan modulations to be performed as a result of the smaller spot size at the pyramid tip focal plane. For additional figures and analysis of our ADC designs, see Kopon et al, 2008.

The criteria used to evaluate relative performance of various designs is the total rms spot size relative to the spot centroid for six different wavelengths spanning the 0.5-1.0μm range in increments of 0.1μm. The Zemax "Atmospheric" surface was used to simulate the atmospheric dispersion with estimated Magellan site parameters (humidity, temperature, etc.).

### 3.1 The 2-Doublet Design

Most ADCs designed and built to date consist of two identical counter-rotating prism doublets (often referred to as Amici prisms) made of a crown and flint glass. The indices of the two glasses are matched as closely as possible in order to avoid steering the beam away from its incident direction. The wedge angles and glasses of the prisms are chosen to correct primary chromatic aberration at the most extreme zenith angle. By then rotating the two doublets relative to each other, an arbitrary amount of first-order chromatic aberration can be added to the beam to exactly cancel the dispersive effects of the atmosphere at a given zenith angle. The 2-Doublet design evaluated in this paper is that of the Arcetri group, which will be used on the LBT WFS and was designed to be diffraction limited over the band 0.6-0.9 μm out to ~65° zenith angle.

The 2-Doublet design corrects the atmospheric dispersion so that the longest and shortest wavelengths overlap each other, thereby correcting the primary chromatism. Secondary chromatism is not corrected and is the dominant source of error at higher zenith angles. To correct higher orders of chromatism, more glasses and thereby more degrees of freedom are needed.

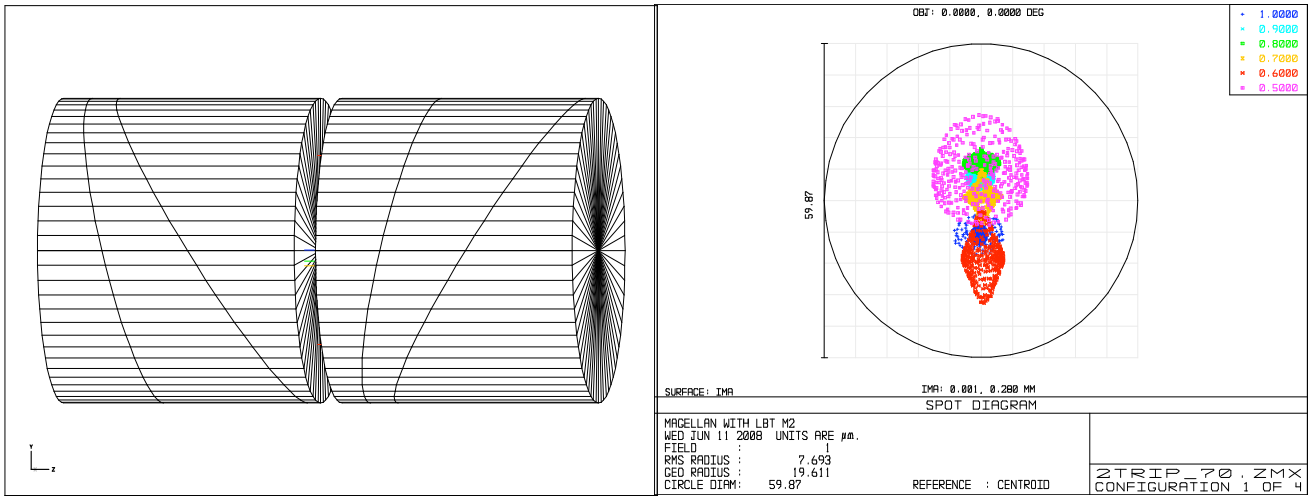

**Figure 5**: Our 2-triplet design at 50 deg zenith (1.55 airmasses). Both primary and secondary chromatism are well corrected, leaving only small amounts of higher order lateral and axial chromatic aberrations. This is the baseline design for the Magellan AO system.

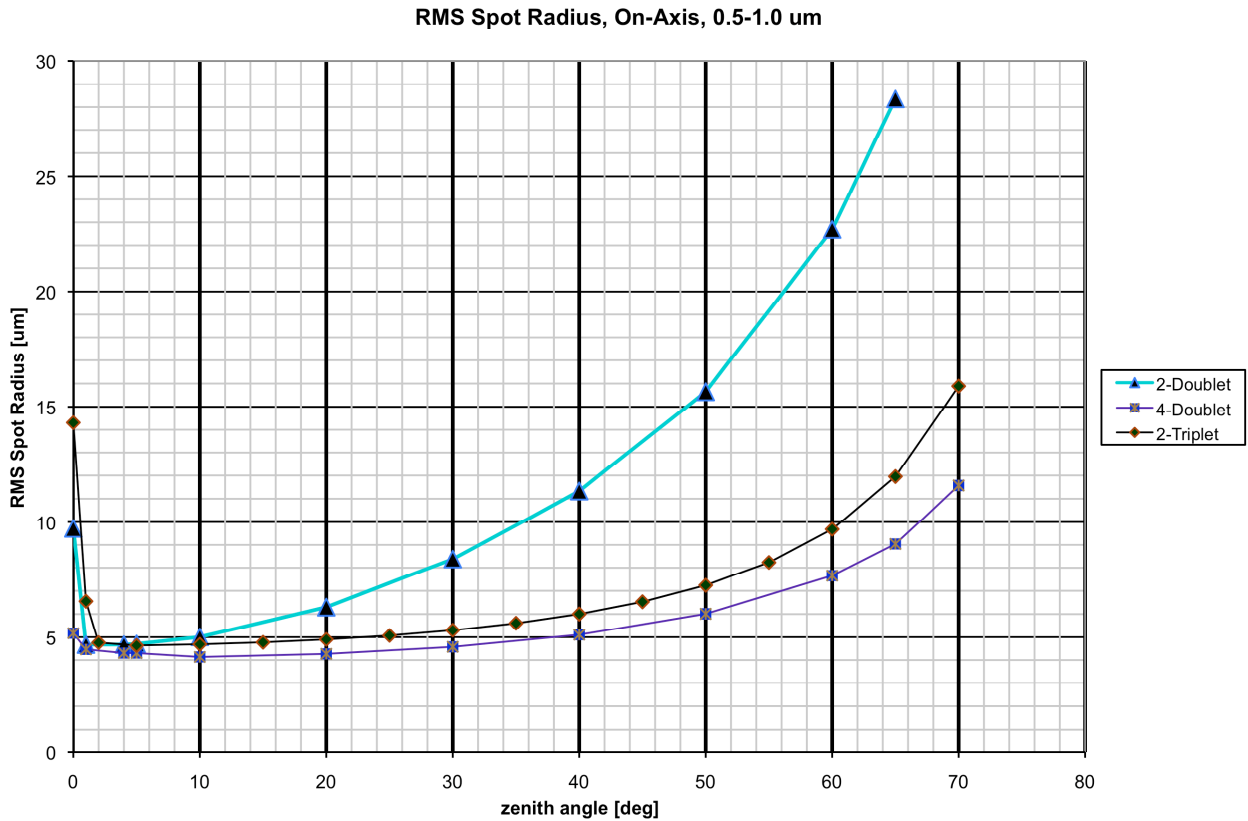

**Figure 6:** RMS spot size vs. zenith angle for the three ADC designs. The 2-triplet ADC is the baseline design for the Magellan AO system due to its relative simplicity and diffraction limited performance down to three airmasses.

### 3.2 The 2-Triplet Design

In our 2-triplet design (Figure 5), a third glass with anomalous dispersion characteristics (Schott's N-KZFS4) is added to the crown/flint pair. Like the doublet, the index of the anomalous dispersion glass was matched as closely as possible to that of the crown and flint. The Zemax atmospheric surface was set to 70 deg zenith and the relative angles of the ADC were set to 180 deg. The wedge angles of the three prisms in the triplet were then optimized to correct both primary and secondary chromatism.

### 3.3 The 4-Doublet Design

This design philosophy can be carried one step further by adding a forth glass and another rotational degree of freedom in a 4-doublet design. The first and second doublets are identical to each other and the third and forth are identical. A multi-configuration Zemax model was optimized for several hours over the full Zenith range, glass catalogue, and possible wedge angles to generate a design that has better overall performance than either the 2-doublet or 3-triplet design. It was also initially thought that the 4-doublet design would be more impervious to errors in fabrication or in the Zemax model atmosphere than the 2-triplet design, since the two doublet pairs can essentially correct primary and secondary chromatism independently of each other. In contrast, the relative amounts of primary and secondary chromatic correction are fixed by the fabricated wedge angles in the 2-triplet design. However, a tolerance analysis shows that the 2-triplet design can handle small fabrication errors and reasonable chromatic fluctuations in the atmosphere due to changes in humidity, temperature, or pressure. Since the 2-triplet design is less complex than the 4-doublet design and has sufficient performance for our needs, it is the baseline design for the Magellan Vis AO system.

## 4. CAOS VISIBLE STREHL SIMULATIONS: HIGH TEMPORAL VARIABILITY

We have been performing simulations of the Magellan AO system using the Code for Adaptive Optics Simulation (CAOS) (Carbillet et al. 2005), an IDL based simulation environment designed specifically for adaptive optics. As a guide for many system parameters we used the simulations conducted by the Arcetri group for the LBT (Carbillet et al. 2003). Our simulated atmosphere is a 6-layer model derived from GMT site surveys at Las Campanas (Lloyd-Hart, 2009). The Magellan AO system will employ modal control, with a theoretical maximum of 585 modes. We have obtained a set of 672 modes calculated from influence functions for the LBT mirror (Brusa, 2009). There will be some differences between these modes and the actual modes achieved at Magellan; however we expect these to have only a modest effect on the following predictions.

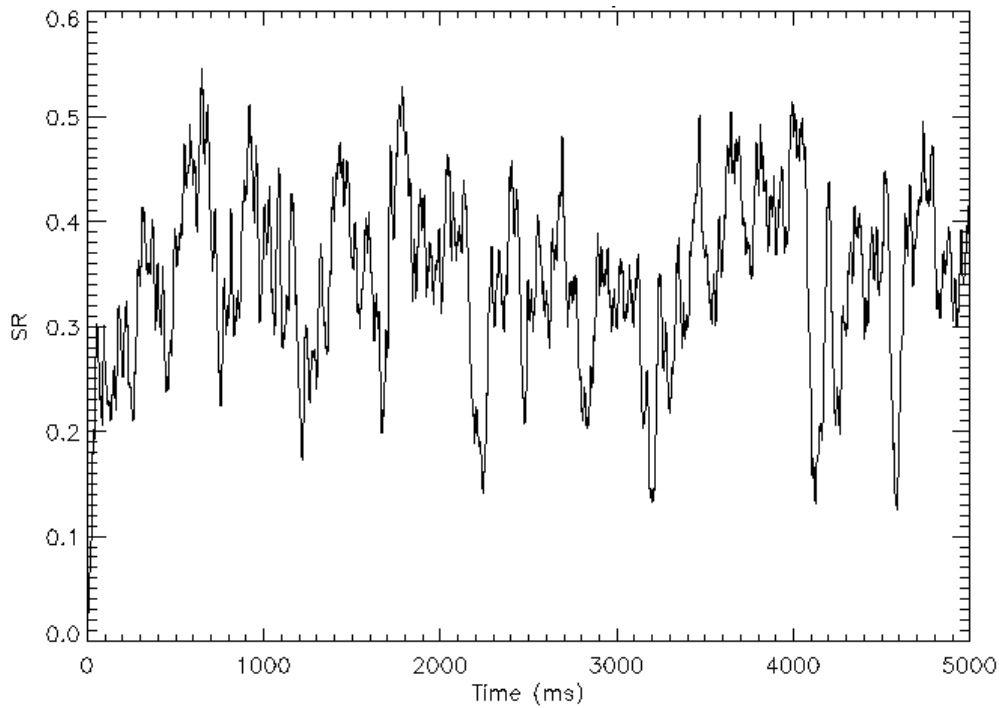

**Figure 7:** A simulated time series of SR at 0.7μm with the LCO atmosphere on a bright guide star. Note the rapid variation in the SR. This simulation motivates the need for a fast asynchronous shutter to block the low Strehl light, thereby optimizing our visible science performance.

We calculate Strehl ratio (SR) from K and R band images saved at 1 ms intervals. We have access to the simulated wavefronts in CAOS, and can calculate SR from the spatial rms wavefront deformation. These are used to crosscheck the results of calculating SR from simulated images and the two methods produce nearly identical results. A 60 nm rms wavefront error is added in quadrature to our predictions to account for static mirror error. This number is based on the 'baseline' specifications used for the LBT672 system. The performance of Magellan AO system as simulated in CAOS is consistent with the predictions of simulations conducted for the very similar LBT system.

A key feature of all of our simulations to date has been the high temporal variability of the short wavelength SR. Figure 7 shows the results of a single simulation run, lasting for 5 seconds with data recorded at 1 ms intervals. For short periods of time the 0.7 μm SR is near 0.5; however, SR also becomes quite poor for significant periods of time.

We expect this variation in SR at visible wavelengths on such short time scales to have significant impact on science operation. The ability to read the CCD47 at fast enough rates will be important, as the shorter exposures will take advantage of the peaks in SR. This would facilitate "Lucky Imaging" style data reduction. If SR can be estimated in real-time, the camera could be shuttered whenever SR drops below a certain value. A shutter can also be used to limit exposure time, even if readouts occur at a slower rate (e.g. 5 ms shutter limited exposures read out in 50 ms (20Hz) to minimize read noise) to aid the "Lucky Imaging" process.

## 5. HIGH SPEED ASYNCHRONOUS SHUTTER: BLOCKING LOW-STREHL LIGHT

The behavior illustrated in Figure 7 will have a detrimental effect on the detection of faint companions due to the expansion of the guide star halo at lower SR. The halo flux at a particular location goes as (1-SR), so the photon noise likewise goes as $(1-SR)^{1/2}$. In an AO corrected image, an additional noise source in the form of coherent speckles will be present. This noise source is proportional to (1-SR) (Racine et al. 1999). So, as SR drops to the lower values evident in Figure 7, detecting a faint companion in the noise due to the halo becomes more and more challenging. Employing the Lucky Imaging technique (Law et al. 2009) to select images with high SR adds a further noise source in the form of detector readout noise.

Using a fast shutter, combined with the ability to estimate SR in real time, it will be possible to reject the periods of low SR without incurring the read noise penalty of Lucky Imaging. We have chosen the Uniblitz model VS-25, customized to be open when powered off. The response time specification is 6ms for the VS-25, with a minimum exposure of 3ms, a repetition rate of 10 Hz continuously and 40 Hz for short periods. Because we will be liquid cooling our VS-25, we expect to exceed these limits. In our initial bench tests we have achieved response times of approximately 5ms, slightly better than the manufacturer specification. Mechanical vibration does not appear to be significant.

The most challenging aspect of our fast shutter low-SR rejection technique will be estimating and predicting SR in real-time. Much effort has been made towards using AO loop telemetry to calculate the system PSF (Veran et al. 2007; Roggemann and Meinhardt, 1993). Our application is slightly less demanding in that we desire only a single parameter rather than the entire PSF. However, this is more than compensated by the need to predict SR about 6 ms in the future. Here we present very early results from an initial attempt using a machine learning algorithm called a Support Vector Machine (SVM). The SVM is a form of linear classifier, meaning it can be used to split data into 2 groups (such as GOOD and BAD). A neural network learning machine is another well known linear classifier, and in fact a neural network can be represented as a type of SVM. Complete description of the technique is beyond the scope of this paper, however an excellent introduction can be found in Burges, 1998.

Determining the PSF without reading out the CCD is calculationally demanding. We also desire to predict SR a small time in the future, so that the shutter control circuit can actuate in *anticipation* rather than *reaction*. The SVM provides a method to utilize a set of parameters that are related to SR, without requiring that their relationship be known *a priori*.

To implement the SVM, we used simulated AO loop telemetry derived from CAOS, including: x and y slopes (slope computer output), pyramid sensor CCD39 frames, and mirror mode amplitudes (commands from the reconstructor). These were chosen because of their possible relationship to SR and the variation of SR with time. They are also available in our wavefront sensor control software. It is also possible that the classifications can be made on metrics other than just SR, such as tip/tilt (measured at the image plane in training and using the command amplitudes for modes 1 and 2). In this case a good image would have both a high SR and a minimum deviation of the centroid position.

During the 'training phase', the SVM algorithm was supplied with telemetry at 3 previous times, along with the known value of SR at t=0 determined from the CCD47. For this initial effort, the training phase lasted 1950 ms, after which the SVM algorithm had determined a model with which to classify SR as good or bad, based on a user supplied threshold (e.g. 0.35). Next, during the 'testing phase' – which would correspond to science data acquisition – the SVM was supplied with only the telemetry parameters and was used to predict the classification of SR 5 ms in the future. Figure 8 shows these first results over 2 seconds. Any time the shutter lines reach the threshold level (0.4) the SVM would have opened the shutter. When at 0, the shutter would be closed. The goal is to maximize the time when the shutter was open with SR above the threshold while minimizing the time the shutter was open below.

Figure 9 shows the SR achieved using the simulated shutter and data, comparing an ideal classifier (perfect knowledge) vs. our SVM performance. We also consider the fraction of time the CCD47 is exposed for any given amount of telescope time (Figure 10). There is clearly a tradeoff in SR vs. telescope time, with signal to noise ratio being the important metric. Future work will analyze the benefits of this technique relative to maximizing signal to noise ratio for faint companions.

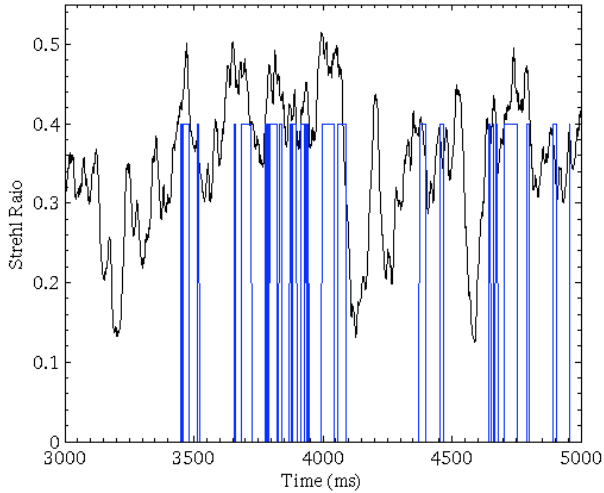
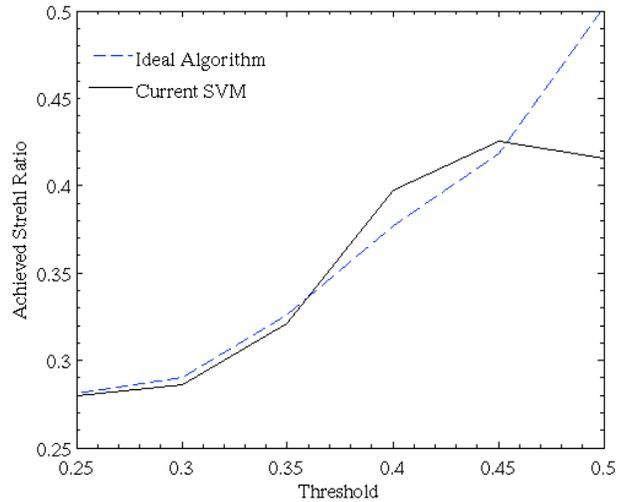

**Figure 8:** Simulated SR at 0.7 μm, with SVM prediction based shutter position overlaid. The SR = 0.4 threshold (blue) line shows the shutter opening mostly while SR is above the threshold. Though it is not open for every possible point above 0.4, it does a good job of rejecting the low SR periods of time, such as the period prior to 3400 ms.

**Figure 9**: Achieved SR vs. classification threshold, from CAOS simulations of the Magellan VisAO system, showing how our current SVM implementation compares to the ideal algorithm

**Figure 10**: Shutter duty cycle. This plot shows the fraction of telescope time that the CCD47 would be exposed vs. threshold SR. As the astronomer gets greedy and pushes for SR > 0.45, his or her duty cycle goes to 0.

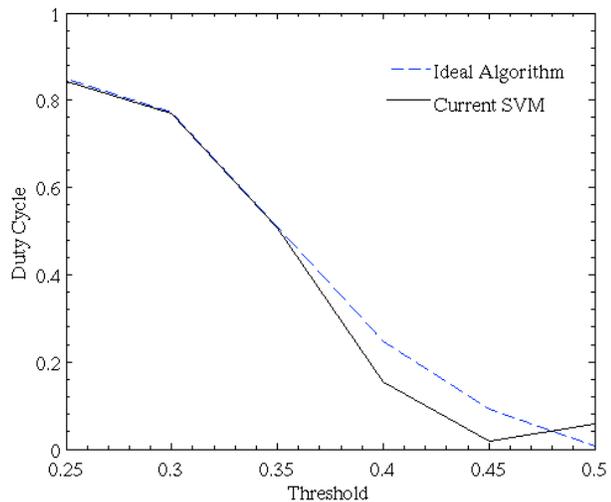

## 6. THE CALIBRATION RETURN OPTIC (CRO) TEST

Because the ASM is a concave ellipsoid, its surface can be tested by placing an interferometer at the far focus and a retro reflector, which we call the calibration return optic (CRO), at the near focus. The CRO, designed by the Arcetri group, is an F/1 parabola with a retro reflecting flat in a double-pass configuration. This interferometric measurement can be used to calibrate the interaction matrix of the system and determine the best actuator settings that "flatten" the mirror (i.e. make it into as perfect an ellipsoid as possible). The CRO test can also be used with a white-light fiber point source at the far focus and the pyramid wavefront sensor, instead of the interferometer, in order to run the ASM and wavefront sensor in closed loop (without the need for natural star light).

This test will be performed initially during system calibration at the solar test tower at the Arcetri Observatory, just as it is being performed on the two LBT ASMs. We also plan to retain the ability to performing this test on the Magellan telescope, particularly during commissioning. In the Arcetri tower, the CRO is stationary w.r.t. the tower and the ASM is aligned to the stationary CRO using a hexapod identical to the hexapod used on the LBT. Because we will not have a hexapod at Magellan that can move the ASM relative to the CRO, it is necessary to build in the added functionality of moving the CRO remotely in order to align it to the ASM. We plan to fabricate a structure that will attach to the windscreen surfaces on the ASM that can hold the CRO at the near focus. The structure will also support a 5-axis (XYZ, θX, θY) remotely controlled fine adjustment stage that can move the CRO relative to the ASM. The CRO and its stage will be repeatably kinematically removable in order to allow us to easily switch back and forth between the CRO test and an on-sky guide star during commissioning.

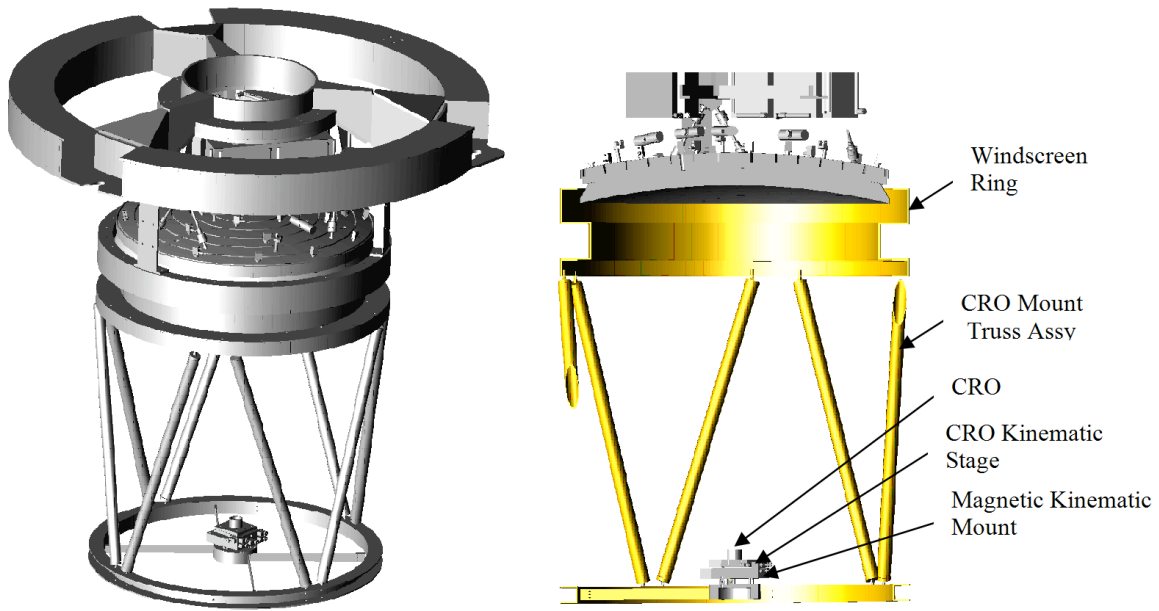

**Figure 11:** Magellan CRO test concept. The removable CRO mount truss will mount to the windscreen. The CRO will be attached to a remotely controlled fine-adjustment stage that can be removed to allow the option of either the CRO test or an on-sky test during commissioning (without removing the truss). The truss will be removed once the commissioning phase concludes.

## 7. MOVABLE GUIDER PROBE: WIDE-FIELD AND SHACK-HARTMANN GUIDING

A movable guider probe with both a wide-field mode and a Shack-Hartmann mode will be used for acquisition, guiding, and telescope alignment/collimation (active optics) as needed before the AO loop is closed. The guider probe design is essentially the same as the conventional Magellan design used on other non-IMACS ports (Schechter et al. 2002). We have changed the focal length of the collimating lens of the Shack-Hartmann E2V CCD to accept the F/16 beam of our secondary mirror, instead of the standard F/11 beam. The layout of our guider ring with the positions of the W-unit and the guider probe is shown in Figure 12.

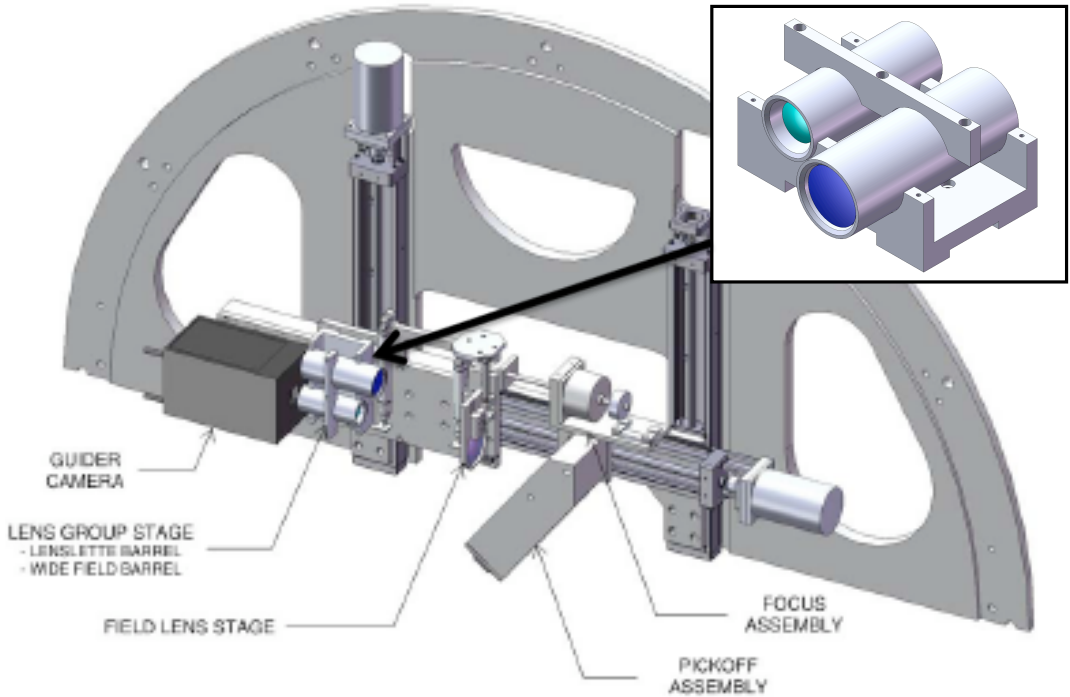

**Figure 12:** Layout of the guider ring assembly. Light from the telescope is reflected by the pickoff mirror and then travels to the focus assembly, the field lens stage, and lens group stage before landing on the CCD. The pneumatic slides allow the selection of either the wide-field acquisition mode, or the SH guiding/collimation mode. Inset: the SH lenslet and the wide-field lens assembly.

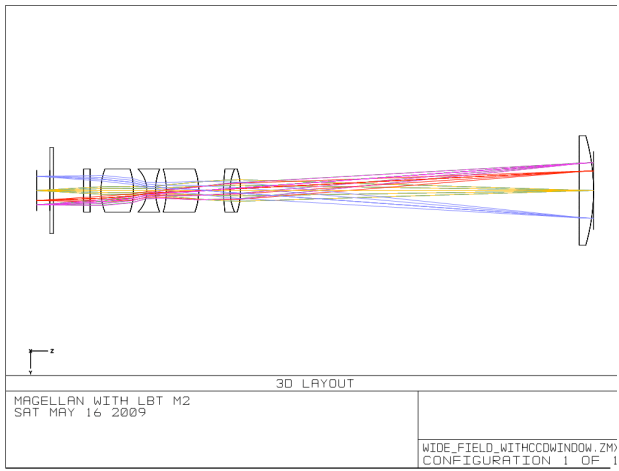

**Figure 13:** Our new custom 50" FOV acquisition camera. Light enters from the right through the field lens (the field lens is on a separate mechanical stage) at the F/16.16 focal plane. The four smaller 20mm dia. lenses and filter are mounted in one lens tube on a pneumatic stage that can be toggled back and forth between the imaging mode and the Shack-Hartmann mode. The filter and CCD window are also shown.

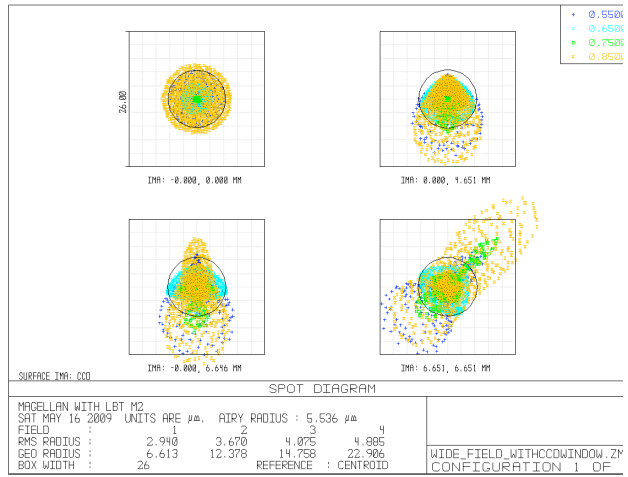

**Figure 14:** Spot diagram for the wide-field lens over the band 550-850 nm. From top left to bottom right: on-axis, 7/10 field point, CCD edge (25 arcsec from on-axis), and CCD corner. The black circle is the 550 nm diffraction limit. The black square is 1/10 of an arcsec on a side.

We have also designed a new custom 50" FOV acquisition camera that can be placed in the center of the field. The design residual of the new wide-field lens is significantly better than seeing limited (0.1") over its 50" field. The wide-field lens will be in a tube that is on a mechanical slide that can move in front of the CCD when wide-field acquisition mode is desired. The lens group operates at F/8.25 (giving 0.05"/pix on each 13μm pixel, but really 0.10"/pix since 2x2 binning is standard) and is well-corrected over the band 550-850 nm. The end of the tube holds a filter that limits the transmitted light to this spectral range. The pneumatic slides allow us to switch easily from the wide-field lens to the SH mode and back. The spot diagram in Fig. 14 shows the quality of the wide field camera design. The black square is the size of one of the Magellan 1k x 1k E2V CCD pixels (0.1" x 0.1" when used in standard 2x2 binning). This design residual is far better than even the best (~0.25") optical seeing conditions that are possible at the telescope.

## 8. CONCLUSION

In this paper we have discussed some of the technical challenges facing all visible adaptive optics systems on large telescopes and our plans to mitigate them on the Magellan VisAO system. Our high actuator density and pyramid wavefront sensor will allow sufficient pupil sampling to correct for the short spatial turbulence scale. Our triplet ADC design will correct for large amounts of atmospheric dispersion at visible wavelengths. Our fast asynchronous shutter will allow us to block the low-Strehl light that is a function of the rapid Strehl variation at visible wavelengths, as predicted by our models. We also present our off-sky retro reflecting calibration test and our new guider camera design that has both a 50" imaging mode and a Shack-Hartmann guider/collimator mode.

## ACKNOWLEDGEMENTS

This project owes a debt of gratitude to our partners and collaborators. The ASM and WFS could not have been possible without the design work of Microgate and ADS in Italy as well as Arcetri Observatory and the LBT observatory. We would like to thank the NSF MRI and TSIP programs for generous support of this project in addition to the Magellan observatory staff and the Carnegie Institute. We would also like to thank Simone Esposito, Andrea Tozzi, and Phil Hinz for providing the Zemax design of the LBT pyramid WFS and Tyson Hare for providing views of the guider ring.